# Extreme Events related with spatial patterns in an all-solid-state laser with saturable absorber

Carlos Bonazzola[1], Alejandro Hnilo[1], Marcelo Kovalsky[1] and Jorge Tredicce[2,3].

[1]*Centro de Investigaciones en Láseres y Aplicaciones (CEILAP), Instituto de Investigaciones Científicas y Técnicas para la Defensa (CITEDEF), Consejo Nacional de Investigaciones Científicas y Tecnológicas (CONICET), J.B. de La Salle 4397, (1603) Villa Martelli, Argentina.*
[2]*Departamento de Física, Facultad de Ciencias Exactas y Naturales, Universidad de Buenos Aires, Intendente Güiraldes 2160, Ciudad Autónoma de Buenos Aires, Argentina.*
[3]*Universite de la Nouvelle Calédonie, ISEA, BP R4, Noumea, Nouvelle Calédonie, France.*
emails: ahnilo@citedef.gob.ar, alex.hnilo@gmail.com

June 30th, 2017.

The passively Q-switched, self-pulsing all-solid-state laser is a device of widespread use in many applications. Depending on the condition of saturation, which is easy to adjust, different dynamical phenomena are observed: continuous wave emission, stable oscillations, period doubling bifurcations, chaos and, in some chaotic regimes, extreme events in the form of pulses of extraordinary intensity. These pulses are also sometimes called "dissipative optical rogue waves". The mechanism of their formation is still unknown. Here, we report the direct observation of the pulse-to-pulse evolution of the transverse pattern with an ultrafast camera (up to $6 \times 10^4$ frames per second). A specific pattern is correlated with the pulse intensity in the periodical regimes. In the chaotic regimes, the extreme events are correlated with some patterns. The series of patterns before and after an extreme event are often the same. These observations demonstrate that extreme events in this system are the consequence of the deterministic nonlinear interaction of few modes. This information plays a crucial role in the development of a theoretical model able to describe the mechanism giving raise to extreme events. The model is expected to lead to the control of their formation at will, what is of practical interest.

## 1. Introduction.

Waves of very high amplitude randomly appearing in deep ocean waters are important phenomena, which received the name of "freak", or "rogue", waves [1]. In the last decade, scientific interest increased on analogous rare or extreme events (EEs) of large amplitude observed in other areas. In Optics, "optical rogue waves" were first observed in the intensity fluctuations of light at the edge of the spectrum produced by ultrashort laser pulse pumped optical fibers in the threshold of super-continuum generation [2,3]. Conditions for their formation were later studied in experiments using optical fibers [4]. Optical EEs were also observed in a VCSEL with an injected signal [5], in fiber lasers [6-9], in Kerr-lens mode locked Ti:Sapphire lasers [10] and, what is our scope here, in self-pulsing (or passive Q-switch) all-solid state lasers with a "slow" saturable absorber (SA) [11]. Reference 12 is a recent review on optical rogue waves.

The all-solid-state passively Q-switched lasers are small, robust, efficient and non-expensive devices. They are used in many applications, from sparks in inner-combustion engines to rangefinders and target illuminators. EEs are easily observed in the chaotic regime of these lasers if the Fresnel number of the cavity is large. This condition is reached easier than not, because the pump diodes are difficult to focus, what naturally produces an active region (and hence, a limiting aperture) broader than the fundamental mode size of the cavity. A laser of this type in the regime with EEs is hence easy to build and robust to operate. Controlling the formation of EEs in this system is hence of practical interest. It would allow the generation, at selected times, of pulses of higher intensity than what is normally obtained from the device. This would be especially useful for laser rangefinders aboard small, low cost unmanned flying vehicles.

Besides, the system is of academic interest to study under favorable conditions the general phenomenon of the formation of EEs: transients disappear instantaneously at the human's timescale, and the parameters are easy to control. The many-mode laser with SA in the limit of small nonlinearity is described by a non-linear Schrödinger equation, hence relating it with other systems where rogue phenomena have been observed. However, when Q-switching is involved the nonlinearity is not small, but dominant. Hence, the theoretical framework to describe the formation of EEs in this system is still to be determined. Our aim is to make observations to guide (and to serve as a test of) that theoretical description.

A key question is whether the number of transverse modes playing a significant role in the dynamics is large or small. If it is large, the theoretical description should involve a transversal Laplacian, as in the broad area VCSEL with SA [13]. This leads to a system of partial differential equations. If the number of modes is small instead, a simpler system of ordinary differential equations may suffice [14]. Previous observations [11,15] suggested that the regimes with EEs in this laser corresponded to the second case. In order to confirm this hypothesis, it was proposed to observe the pulse-to-pulse evolution of the transverse pattern of the laser spot. The steady pattern observed with a standard CCD camera (50 frames per second, fps) was found to be quite complex, but that pattern was the average over hundreds of pulses, because of the high repetition rate (tens of KHz) of the laser.

In this paper we present the results of the observation of the pulse-to-pulse evolution of the transverse pattern of an all-solid-state self-pulsing (or passively Q-switched) laser. This is possible thanks to the use of an ultrafast camera (up to $6\times10^4$ fps). The purpose of this observation is to guide the theoretical modeling by outlining the key points to be predicted. Note that in most experimental work on nonlinear dynamics of optical systems the basic mathematical description of the problem is known. Here instead, we start from an existing system where nonlinear phenomena (the EEs) arise spontaneously, and our goal is to find a theoretical description that allows us to control the formation of these phenomena for a practical application.

At this stage of the research, the questions are: *(i)* is the complex transverse pattern observed with the standard (50 fps) camera the superposition of many different patterns, or it is a single pattern that repeats itself in all the pulses in the chaotic series?; *(ii)* in the case the patterns change from one pulse to the next, how many different patterns exist?; *(iii)* are the EEs associated with a single pattern?; *(iv)* an EE is caused by the appearance of a "hot spot" of high intensity, or by a uniform high illumination of the whole pattern?; *v)* are the EEs predictable? F.ex.: do the patterns before and after an EE follow a particular sequence (as it was observed in the pulse intensities [15]) or not?

In the next pages, the answers to these questions are presented. The key points a satisfactory theoretical description must reproduce are summarized.

## 2. Experimental setup.

The setup is shown in the Fig.1. The output of a 2W (@808 nm) CW laser diode is focused by a gradient index (GRIN) lens to a spot 0.8 mm diameter into a Nd:YVO$_4$ crystal, 1% doped with standard dichroic coating. The V-shaped laser cavity has a folding HR concave mirror (radius=10 cm) and a plane output coupler (reflectivity= 98%). The operating wavelength of the laser is 1064 nm, linearly polarized. The mode size varies strongly between mirrors, with a waist near the output coupler. The cavity's Fresnel number is $\approx$10. A solid-state SA (Cr:YAG, 90% transmission unbleached) is placed between the folding mirror and the output coupler at position X. Adjusting this position, the mode size at the SA changes and hence the condition of saturation. This is the main control parameter in this system. As X is varied different dynamical regimes appear, from stable Q-switch to periodic bifurcations and chaotic regimes with and without EEs. The average output power is practically the same in all the regimes.

A beam splitter divides the laser output in two: one part is focused into a pin fast photodiode (100 ps risetime) connected to a digital storage oscilloscope (PicoScope 6403B: 350 MHz bandwidth, 5 GS/s, memory 1 GS); the other part is projected in a translucent screen and the image of the laser spot is observed and recorded with an ultrafast camera (Photron Fastcam SA-3). The oscilloscope records the shape of the laser pulses, while the camera simultaneously records the spot pattern. The synchronous start of both recordings is ensured by blocking (and then unblocking) the laser beam between the output mirror and the beam splitter (see Fig.1), and by setting the oscilloscope and the camera in "auto trigger" mode.

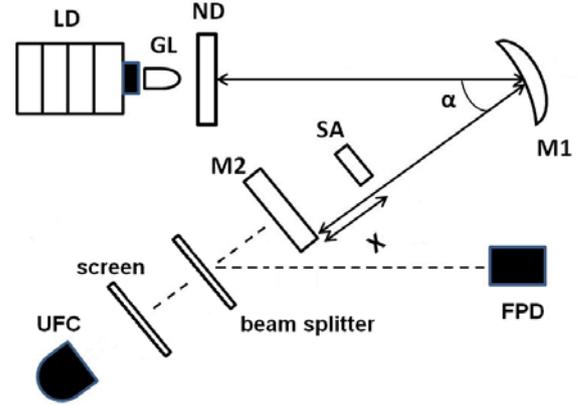

Figure 1: Sketch of the setup. LD: pump laser diode, 2 W cw at 808 nm; GL: GRIN lens; ND: Nd:YVO$_4$ slab 3×3×1 mm$^3$ 1% doped; M1: folding mirror (HR, R = 10 cm); M2: output mirror (reflectivity 98%, plane); SA: Cr:YAG crystal, transmission (unbleached) 90%; $\alpha$ = 20°; distance ND-M1=13 cm; distance M1-M2 = 7 cm; the position X is adjusted to get different dynamical regimes; FPD: fast photodiode connected to a digital memory oscilloscope; UFC: ultrafast camera recording single pulse spot images; the screen is translucent. By unblocking the laser beam between M2 and the beam splitter, the camera and the oscilloscope, which are set in the "auto trigger" mode, start recording at the same time.

The spatial resolution of the camera decreases if the number of fps is increased. There is a discrete menu of "fps × resolution" combinations allowed. Even in the stable periodic regimes, the repetition rate of the laser does not fit exactly the values in this menu and besides, there is a "blind time" after each camera shot. For these reasons, during a long run the patterns corresponding to some pulses are lost (an illustration of this effect is shown in Fig.2). In the chaotic regimes the situation is even worse, for the pulse separation changes wildly. Some patterns may be then missed, or some camera shots may record superimposed the patterns corresponding to two successive pulses that are unusually close in time. Fortunately, if the fps of the camera and the average repetition rate of the laser roughly fit, and some special care is taken (see the next Sections) the number of pulses missed or jammed in a long series is found to be low. In all the runs discussed in this paper, the camera is set to 25000 fps, the spatial resolution is 128×80 pixels and the blind time is 1.14μs. Be aware that the camera records the time-integrated pattern of each pulse. There is no camera fast enough to record the pattern evolution during a ≈100 ns pulse.

A brief note on the definition of EE: it usually is: *a)* amplitude higher than twice the "significant wave height" or "significant intensity" $I_{1/3}$, which is the average calculated among the set of the 1/3 highest events in the series. Alternatively: *b)* amplitude higher than 4 times the standard deviation. Here we use the definition *b)*. The kurtosis K is an additional measure of non-Gaussian feature, K>3 means a distribution with a tail longer and higher than that of a Gaussian one.

## 3. Observations.
*3.1 An example: period three regime.*

Depending on the position X of the SA, different dynamical regimes are observed. The intensity output and the sequence of spatial patterns for a period-3 regime are displayed in the Fig.2 (laser repetition rate ≈23 KHz). It is clearly seen that the periodic evolution of the intensity corresponds to a periodic repetition of spot patterns. This result was expected but, as far as we know, this is the first time it is actually observed. The pulse with the highest intensity corresponds to a 4-fold pattern (called "E", see next section), while the two lower ones to $TEM_{00}$-like patterns (called "A"). The sequence repeats indefinitely.

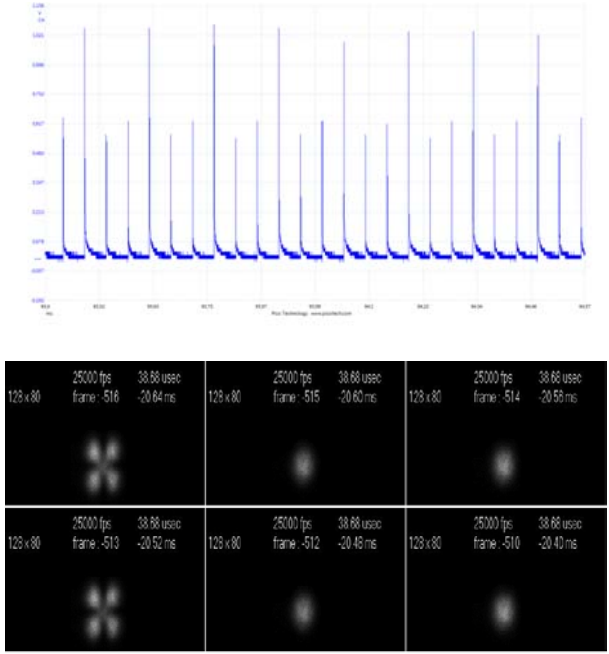

Figure 2: Pulse intensity and associated patterns for a period-3 regime. Note the successive labels of the camera shots (from -516 to -510). As an illustration of a typical error, the shot -511 is between two successive pulses and is an empty image, not shown.

*3.2 Chaotic regime with EEs.*

We then adjusted the SA position so that the laser entered into a chaotic regime. The time separation between pulses changes in an irregular way (average repetition rate ≈27 KHz) and the images corresponding to some pulses can be missed, or superimposed. Besides, there is a slight drift between the internal clocks of the oscilloscope and the camera. For these reasons, the identification of the pattern corresponding to each pulse is not immediate. We develop a computer code, based on maximizing the correlation between two series: the integral below the pulses in the oscilloscope trace, and the integrated intensities in the camera images. We find in this way that the clocks in the oscilloscope and the camera have a relative drift ≈10 μs/s. We check the results provided by the code by visual inspection over short sections of the series and find them practically perfect.

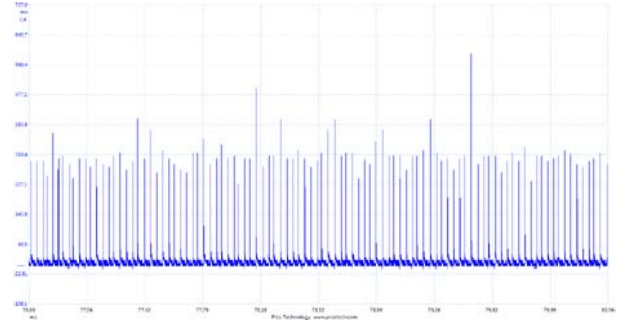

Figure 3: Zoom of the oscilloscope trace for run 2b9/29/16. This chaotic series has $d_E = 8$ and two positive Lyapunov exponents. Note the EE between t=79.26 and 79.62. The complete series has 27779 pulses, recorded in one second in real time.

We are interested in chaotic regimes displaying EEs. In what follows, we discuss the results of a particular run (internal identification: 2b9/29/16, see Fig.3). It is representative of the results obtained in many other experimental runs. If the average peak intensity is scaled to 100, the EE threshold is 161 (Fig.4), and 64 pulses are EEs. The value of $d_E$ is measured by the decay of the fraction of false neighbors, and is between 8 and 10. Two Lyapunov exponents are positive, and K=5.1. The time series are analyzed with the TISEAN free software package.

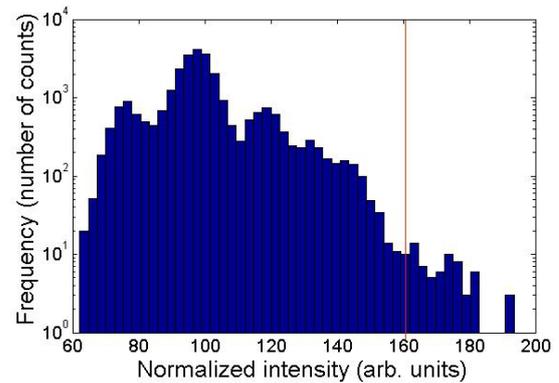

Figure 4: Histogram of peak pulse intensities. The horizontal axis is normalized such that the average intensity =100. The threshold value for EEs is 161 (vertical red line), K = 5.1.

We identify 8 different types of patterns, that we name "A" to "H" (Fig.5). The relative frequency of each pattern in the series is indicated. Now we can answer question *(i)*: the pattern observed at low time resolution is the superposition of hundreds of different, simpler patterns. We can also answer question *(ii)*: there are few types of patterns. Besides, some of the less frequent ones may correspond to a superposition of two: this is the case of type "F" (possible superposition of "C" and "D") and "H" ("C" and "E"). This reduces the number of actual types to 6. Finally, note that the 5 types "A" to "E" suffice to take into account 99% of the pulses.

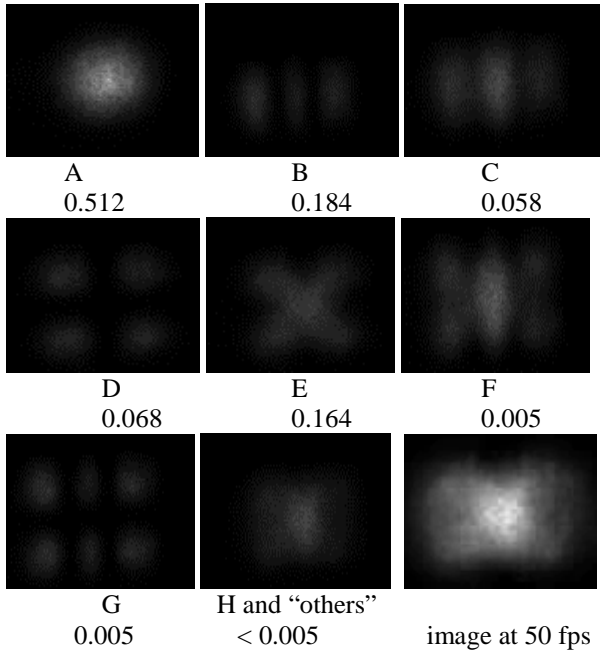

Figure 5: Types of patterns observed. Their relative frequency is indicated. It is possible that F=C+D, and H=C+E. The steady spot observed with a standard camera (50 fps) is shown as a reference.

In the set of 64 pulses above the EE threshold, 36 correspond to the type "E", 23 to "C", 2 to "F", 1 to "H", and 1 is missed (it fell into the blind time of the camera). Now we can answer question *(iii)*: the EEs are not associated with a single pattern. Yet, almost all of them are associated with only two patterns, and one of them is dominant (type "E" amounts to 58% of the EEs). The statistics is too scarce to produce significant histograms of the sets of EEs, but the type "E" ones have a higher scaled average intensity (191) than the "C" (170,7) and the others (171,2). In fact, the type "E" fills the rightmost end of the distribution in Fig.4. Remarkably, there are no EEs of the "A" and "B" types which are, by far, the most frequent of all in the complete series.

We can also answer the question *iv)*: the EEs are not caused by a "hot spot", but instead by a uniform brighter illumination of the whole pattern. This pattern is *not* exclusive of the EEs.

*3.3 Sequences of patterns near an EE.*

Now we consider the question *v)*, which regards the predictability of the EEs. We then study their neighborhood. Since the "C" and "E" sum up more than 90% of the total of EEs, we focus on these types.

A few words on notation: we indicate with a small letter the pattern corresponding to a pulse of normal intensity, and with a capital letter an EE (f.ex., the sequence of patterns in Fig.6 is *eabaEabae*). An asterisk means "any pattern", a dash means a pulse missed by the camera.

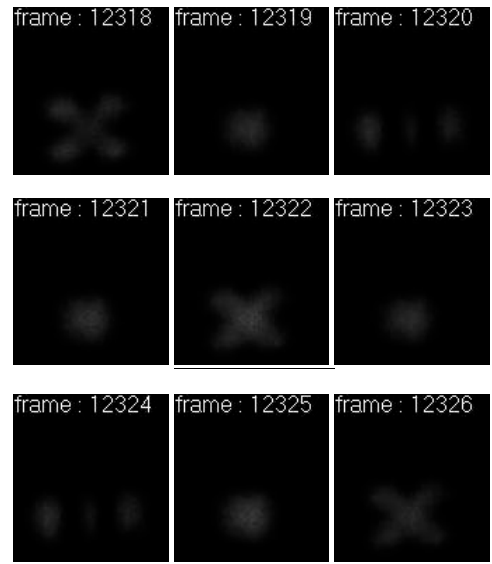

Figure 6: The most frequent sequence of patterns in the set of EEs (14 or 15 out of 64) *eabaEabae* which is, in addition, curiously symmetrical in time. The EE is underlined (frame 12322).

Regarding the "E"-type of EEs, 14 (probably 15, for there is one *eab-Eabae*) out of 36 sequences are the curiously symmetrical in time *eabaEabae* shown in Fig.6. A less restrictive definition, 7-letters sequences, *abaEaba* increases the number only a bit, to 18. The number of sequences *eabaE**** (which are of interest to herald the EE) is 20 so that, within the set of the EEs, *eaba* heralds the appearance of a "E"-type EE with a probability of 0.31. Yet, this is no so if the set of *all* the pulses in the series is considered (see Section 3.4).

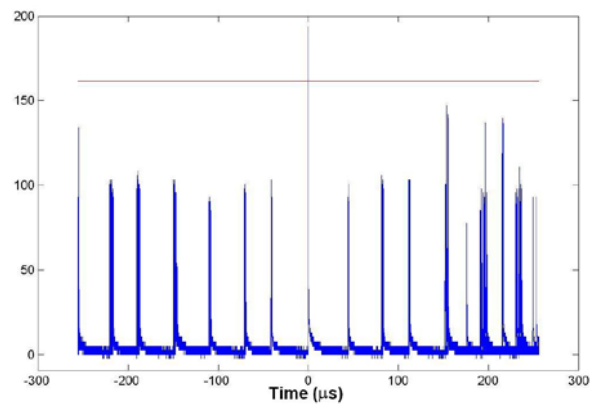

Figure 7: Superposition of the 14 oscilloscope traces of the sequence *eabaEabae*. The horizontal (red) line indicates the EE limit.

A striking example of regularity of pulse intensities and time between pulses in the neighborhood of an EE is displayed in the Fig.7, where the 14 oscilloscope traces corresponding to the sequence *eabaEabae* are plotted superimposed.

Regarding the "C"-type of EEs, and contrarily to the "E"-type, there is no dominant sequence of 9 letters. Instead, some "rare" patterns like *d, f* and *g* appear. The sequence *adaC\*\*\** occurs 14 times, so that within the set of the EEs, *ada* heralds a "C"-type EE with a probability of 0.22. Once again, this is no so in the set of all the pulses (see Section 3.4).

The regularities observed demonstrate that the formation of EEs in this system follows a deterministic mechanism. F.ex., the p-value corresponding to the null hypothesis that the *eabaEabae* sequence appears randomly (in other words: it is the result of an experiment where the probability of appearance of this sequence in each trial is $1,1\times10^{-5}$ and there are 14 results in 64 trials) is $3\times10^{-32}$. This extremely small p-value demonstrates that this sequence is *not* the result of a statistical fluctuation.

*3.4 Statistics on the complete series.*

In the previous section the probabilities of predicting an EE from the sequence of patterns, within the set of EEs, is considered. A question of practical interest is to whether a sequence of patterns exists that heralds the appearance of an EE in the set of *all* the pulses in the series. In order to perform this task, a computer code able to identify each type of pattern (that is: to translate each of the 27779 images into one of the 8 letters in Fig.4) is developed. The error of this code is checked by visual inspection of short subsets of the series. We find that the error in recognizing a pattern strongly depends on the type of the pattern. An average over all the types, weighed by their frequency, indicates an error ≈4%. This number is mainly determined by the relatively high probability of error in recognizing the pattern "A". This code may be useful whenever a pattern recognition task is required.

The number of sequences *eabaeabae* (that is: regardless whether the *e* in the middle is an EE or not) is 321. This means that the sequence appears with a probability ≈1% in the set of all the pulses in the series. It is far smaller than the probability within the set of EEs (15/64 ≈23%). On the other hand, the probability this sequence to happen by chance (calculated from the fractions in Fig.5) is only ≈$10^{-5}$. The same calculus for the *adac* sequence shows that it has a probability to happen by chance of ≈$10^{-3}$, but there are 328 of these sequences in the complete series (>1%), and 14/69≈ 20% within the set of the EEs. In other words: the sequences *eabaeabae* and *adac* appear, in the complete series, much more often than expected if the patterns were random, and they do so even more often within the set of EEs. These results indicate that the system follows some preferred "roads" of patterns in its evolution, and that at least two of these roads are also the most likely to produce an EE. The identification of these roads (actually, orbits within the attractor) is the first step to achieve the management of EEs by control of chaos. The next step is to find some practical way to pick them out to stabilize them at will.

Despite the regularities found, the probability of heralding an EE from a sequence of patterns is poor. The number of sequences *eaba* is 2173, but only 15 EE (all of type "E") follow. The number of sequences *ada* is 1369, but only 14 EE (all of type "C") follow. Therefore, the sequence of patterns does not provide a reliable way to herald an EE. There is a too high number of "false alarms".

Finally, note that practically all EEs belong to the types "E", "C" or "F". These types also have higher average energies in the complete series, by a factor ranging from 1.5 to 2 larger than the other types. The images in the Fig.5 suggest that the cause of this higher efficiency is that these types are the ones best filling the transverse area. It is then possible to hypothesize the basic mechanism of formation of EEs in this way: the EEs are pulses belonging to the spatial distributions most efficient to extract energy from the active medium and that, *in addition*, for some reason still to be explained, extraordinarily succeed in doing it. The reason of this success most probably lies in the spatial distribution of the bleaching of the SA. Recall that EEs are *not* preceded by a long time elapsed since the last pulse (that is: they are not due to the simple accumulation of gain because of a longer pump time). Yet, they are followed by a relatively long "dead" time until the next pulse [15]. Recall also that Cr:YAG is a slow SA, so that it retains memory of the bleaching level reached in the previous pulses. This is the ultimate reason why the dynamics is not trivial.

**4. Conclusions.**

The ultrafast camera reveals that the relatively complex steady transverse pattern observed with a standard CCD (or by the eye) is the superposition of hundreds of simpler patterns, that the number of different types of these simpler patterns is small, that the patterns of the EEs belong to an even smaller set (and that they are free of "hot spots"), and that the sequences of patterns before (and sometimes after) an EE are fairly regular, although they do not allow heralding the EEs with satisfactory efficiency.

Regarding the development of a theoretical description of the formation of EEs in this laser, we can summarize the observations that description must fit, as follows:
1) Chaotic regimes have been observed with all values of $d_E \geq 4$, but no regime with EEs has been observed with $d_E < 6$. EEs are easier to observe with increasing Fresnel number. EEs are observed in regimes with one or more positive Lyapunov exponents as well.
2) The description does not need to include the transversal Laplacian. It suffices to take into account the interaction within a set of relatively few modes that are previously defined, as in [14]. The time series analyzed in this paper needs 5 patterns to take into account 99% of the pulses, but in other recorded series this number is even smaller. The smallest observed number of different patterns in a series with EEs is 3.
3) The regimes with EEs show a decay of the correlation between the time series recorded at different points in the transverse section. The chaotic regimes without EEs do not always show this decay.
4) The patterns associated with EEs are the most efficient to extract energy from the active medium.

5) The time before the appearance of an EE (measured since the previous pulse) is well defined and near the average time between successive pulses. On the contrary, the time after an EE (until the next pulse) is longer than the average, what is most reasonable.

6) The pulse separation and intensity, as well as the sequence of patterns, just before and after an EE, are more regular than for a non-EE pulse. Yet, the inverse is not true: it is *not* possible to predict (with satisfactory accuracy) the appearance of an EE from the observation of a certain sequence of pulses or patterns.

The observed regularities before and after an EE mean that the EEs occur inside a well defined region in phase space, or inside a well defined set of orbits. The formation of EEs can be then managed (in principle, at least) by stabilizing these orbits through algorithms of control of chaos. Regarding the practical goal of controlling the formation of EEs at will, the observation that the orbits leading to an EE are far more frequent than by chance is very important. In principle, the human operator of the laser must wait until one of these orbits appears, and then (if he decides that an EE is necessary at that moment) actuates on the system so that the EE is produced. The probability of occurrence of such orbits in this non-optimized prototype is ≈1% (≈2% if we consider the two main types of EEs). It means an average "waiting period" (until the appropriate orbit appears) below 3 ms in real time. As the time required by a human operator to take a decision is much longer than that, this "waiting period" imposes no practical limitation, even if the efficiency to detect the orbit and to produce the EE were poor.

The precise way the orbit must be perturbed in order to produce an EE will be determined once the details of the dynamics are revealed by the theoretical description. We anticipate the spatial distribution of the bleaching in the SA to play a major role.

**Acknowledgements.**

Many, but really many thanks to Prof. P.Mininni and Prof. P.Covelli, *Laboratorio de Turbulencia Geofísica, Departamento de Física, Facultad de Ciencias Exactas y Naturales, UBA,* for lending us the ultrafast camera. This material is based upon work supported by the Air Force Office of Scientific Research under contract number FA9550-16-6-0045, "Nonlinear dynamics of self-pulsing all-solid-state lasers". It also received support from the grant CONICET PIP11-077 "*Desarrollo de láseres sólidos bombeados por diodos y de algunas de sus aplicaciones*", and the project ECOS-Sud "Extreme Events in Nonlinear systems", A14E03.

**References.**

[1] C.Kharif, E.Pelinovsky and A.Slunyaev, "Rogue waves in the Ocean", Springer-Verlag, Berlin-Heidelberg, 2009.
[2] D.Solli, C.Ropers, P.Koonath and B.Jalali,"Optical extreme events", *Nature* **450**, 1054 (2007).
[3] C.Finot, K.Hammami, J.Fatome, J.Dudley and G.Millot, "Selection of Extreme Events Generated in Raman Fiber Amplifiers Through Spectral Offset Filtering", *IEEE J. Quant. Electron.* **46**, 205 (2010).
[4] A.Montina, U.Bortolozzo, S.Residori and F.T.Arecchi, "Non-Gaussian statistics and extreme waves in a nonlinear optical cavity", *Phys.Rev.Lett.* **103**, 173901 (2009).
[5] C.Bonatto, M.Feyereisen, S.Barland, M.Giudici, C.Masoller, J.Ríos Leite and J.Tredicce; "Deterministic optical extreme events", *Phys.Rev.Lett.* **107**, 053901 (2011).
[6] J.Soto-Crespo, P. Grelu and N.Akhmediev; "Dissipative rogue waves: Extreme pulses generated by passively mode-locked lasers", *Phys.Rev.E* **84**, 016604 (2011).
[7] A.Zaviyalov, O.Egorov, R.Iliew, and F.Lederer; "Rogue waves in mode-locked fiber lasers", *Phys.Rev.A* **85**, 013828 (2012).
[8] C.Lecaplain, P.Grelu, J.Soto-Crespo and N. Akhmediev; "Dissipative Rogue Waves Generated by Chaotic Pulse Bunching in a Mode-Locked Laser", *Phys.Rev.Lett.* **108**, 233901 (2012).
[9] A.Runge, C.Aguergaray, N. Broderick and M. Erkintalo; "Raman rogue waves in a partially mode-locked fiber laser", *Opt.Lett.* **39**, 319 (2014).
[10] M.Kovalsky, A.Hnilo and J.Tredicce, "Extreme events in the Ti:Sapphire laser", *Opt.Lett.* **36**, 4449 (2011).
[11] C.Bonazzola, A.Hnilo, M.Kovalsky and J.Tredicce, "Optical rogue waves in the all-solid-state laser with a saturable absorber: importance of the spatial effects", *J.Opt.* **15**, 064004 (2013).
[12] J.Dudley, F.Dias, M.Erkintalo and G.Genty,; "Instabilities, breathers and rogue waves in Optics", *Nat.Photon.* **8**, 755 (2014).
[13] C.Rimoldi, S.Barland, F.Prati and G.Tissoni, "Spatio-temporal extreme events in a laser with a saturable absorber", *Phys.Rev.A* **95**, 023841 (2017).
[14] J.Dong, K-I Ueda, P.Yang, "Multi pulse oscillation and instabilities in microchip self-Q-switched transverse-mode laser", Opt.Express. 16980 (2009).
[15] C.Bonazzola, A.Hnilo, M.Kovalsky and J.Tredicce, "Features of the extreme events observed in an all-solid-state laser with a saturable absorber", *Phys.Rev.A* **92**, 053816 (2015).